# GRAVITY: a four-telescope beam combiner instrument for the VLTI

S. Gillessen*[a], F. Eisenhauer[a], G. Perrin[b,i], W. Brandner[c], C. Straubmeier[d], K. Perraut[e], A. Amorim[f], M. Schöller[g], C. Araujo-Hauck[d], H. Bartko[a], H. Baumeister[c], J.-P. Berger[e,g], P. Carvas[f], F. Cassaing[h,i], F. Chapron[b,i], E. Choquet[b,i], Y. Clenet[b,i], C. Collin[b,i], A. Eckart[d], P. Fedou[b], S. Fischer[d], E. Gendron[b,i], R. Genzel[a], P. Gitton[g], F. Gonte[g], A. Gräter[a], P. Haguenauer[g], M. Haug[a], X. Haubois[b,i], T. Henning[c], S. Hippler[c], R. Hofmann[a], L. Jocou[e], S. Kellner[a], P. Kervella[b,i], R. Klein[c], N. Kudryavtseva[c], S. Lacour[b,i], V. Lapeyrere[b,i], W. Laun[c], P. Lena[b,i], R. Lenzen[c], J. Lima[f], D. Moch[a], D. Moratschke[d], T. Moulin[e], V. Naranjo[c], U. Neumann[c], A. Nolot[e], T. Paumard[b,i], O. Pfuhl[a], S. Rabien[a], J. Ramos[c], J.M. Rees[b,i], R.-R. Rohloff[c], D. Rouan[b,i], G. Rousset[b,i], A. Sevin[b,i], M. Thiel[a], K. Wagner[c], M. Wiest[d], S. Yazici[d], D. Ziegler[b,i]

[a]Max-Planck-Institute for extraterrestrial physics, Giessenbachstraße, D-85748 Garching, Germany;
[b]LESIA, Observ. de Paris Meudon, 5, place Jules Janssen, 92195 Meudon Cedex, France;
[c]Max-Planck-Institut für Astronomie, Königstuhl 17, 69117 Heidelberg, Germany;
[d]I. Physikalisches Institut, Universität zu Köln, Zülpicher Strasse 77, 50937 Köln, Germany;
[e]Laboratoire d'Astrophysique, Observatoire de Grenoble, BP 53, 38041 Grenoble Cedex 9, France;
[f]SIM, Fac. de Ciências da Univ. de Lisboa, Campo Grande, Edif. C1, P-1749-016 Lisbon, Portugal;
[g]European Southern Observatory, Karl-Schwarzschild-Straße 2, 85748 Garching, Germany;
[h]ONERA/DOTA, 29 avenue de la Division Leclerc, F-92322 Chatillon, France;
[i]Groupement d'Intérêt Scientifique PHASE (Partenariat Haute résolution Angulaire Sol Espace) between ONERA, Observatoire de Paris, CNRS and Université Paris Diderot.

## ABSTRACT

GRAVITY is an adaptive optics assisted Beam Combiner for the second generation VLTI instrumentation. The instrument will provide high-precision narrow-angle astrometry and phase-referenced interferometric imaging in the astronomical K-band for faint objects. We describe the wide range of science that will be tackled with this instrument, highlighting the unique capabilities of the VLTI in combination with GRAVITY. The most prominent goal is to observe highly relativistic motions of matter close to the event horizon of Sgr A*, the massive black hole at center of the Milky Way. We present the preliminary design that fulfils the requirements that follow from the key science drivers: It includes an integrated optics, 4-telescope, dual feed beam combiner operated in a cryogenic vessel; near-infrared wavefront-sensing adaptive optics; fringe-tracking on secondary sources within the field of view of the VLTI and a novel metrology concept. Simulations show that 10 µas astrometry within few minutes is feasible for a source with a magnitude of $m_K = 15$ like Sgr A*, given the availability of suitable phase reference sources ($m_K = 10$). Using the same setup, imaging of $m_K = 18$ stellar sources in the interferometric field of view is possible, assuming a full night of observations and the corresponding UV coverage of the VLTI.

**Keywords:** Interferometry, Near-infrared, Black Hole, Galactic Center, Sgr A*, VLTI

*ste@mpe.mpg.de; phone +49-89-30000-3839; fax +49-89-30000-3390;

# 1. INTRODUCTION

Astrometry is a powerful tool for carrying out fundamental measurements and for unveiling the laws of Nature and the structure of the Universe. The pioneering work from Brahe and Kepler lead to the discovery that the planets revolve the Sun on elliptical orbits, the basis for the Newtonian theory of gravitation. Astrometric parallax measurements unambiguously determine the distances to stars in the solar neighborhood, or even to star forming regions throughout the Milky Way in the case of intercontinental long baseline radio interferometry (VLBI, [1]). VLBI also yielded precision measurements of the dynamics of a warped accretion disk around the super massive black hole (SMBH) in the galaxy NGC 4258 [2]. Speckle and adaptive optics, near-infrared (NIR) astrometry of stars demonstrated that Sgr A*, the compact radio source at the centre of the Milky Way must be a SMBH of about 4 million solar masses [3,4,5].

NIR interferometry with the Very Large Telescope Interferometer (VLTI) will open a new era of high-resolution, narrow-angle precision astrometry. It will make possible phase-referenced imaging of faint sources at mas resolution, and 10 – 100 µas precision astrometry, all at the exquisite sensitivity provided by the large collecting area of the VLT. With the GRAVITY instrument, the angular resolution and astrometric precision will improve over current adaptive optics (AO) imaging on current 10 m class telescopes by a factor of about 10 – 30, and it will be superior even to the next generation of telescopes planned, the ELTs. At the aimed-at level of precision, the motions of many astronomical sources become measurable within a reasonable time frame of a few years.

The aim of GRAVITY is to offer to the observers an instrument that interferometrically combines near-infrared (NIR) light collected by the four unit telescopes of ESO's Very Large Telescope, using adaptive optics at the telescope level and fringe-tracking at the interferometer level. This instrument can be used in an imaging mode, yielding an unprecedented resolution of ~ 3 mas in the NIR for objects that can be as faint as $m_K = 18$ when using a fringe-tracking star of $m_K = 10$. In its astrometric mode, GRAVITY will allow to measure distances between the fringe-tracking star and a science object to an accuracy of 10 µas. At 100 pc, a velocity of 10 µas/yr corresponds to 5 m/s, at 1 Mpc to 50 km/s.

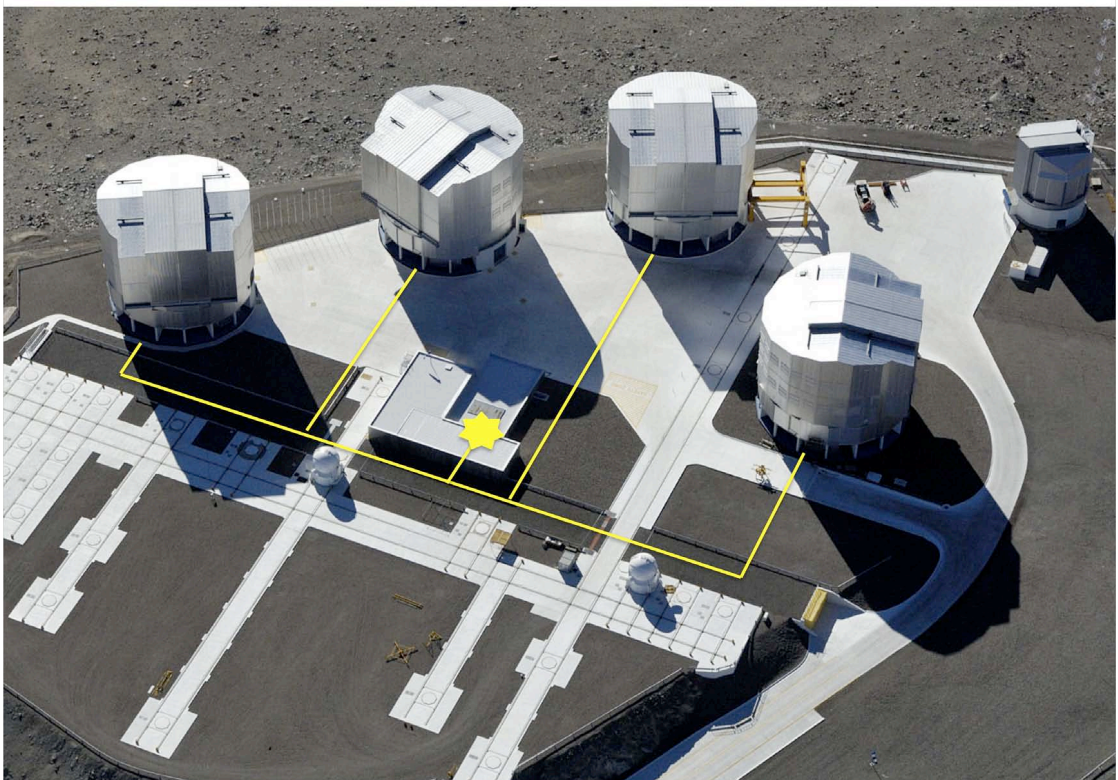

Figure 1: The VLT interferometer at Paranal (Chile), future site for GRAVITY.

## 2. SCIENTIFIC MOTIVATION

### 2.1 Overview

GRAVITY will revolutionize dynamical measurements of a plentitude of celestial sources.

- Using stellar dynamics around Sgr A*, GRAVITY will carry out tests on the equations of motion around a SMBH, a previously unexplored regime of gravity [6]. In particular, the effect of relativistic periastron advance can be tested.

- GRAVITY has a fair chance of detecting the spin of the SMBH by means of its influence on stellar orbits. If combined with a measurement of the quadrupole moment of the SMBH, GRAVITY will even be able to test the 'no-hair' theorem for black holes [7].

- If the current interpretation of the NIR flares from Sgr A* as localized events in the innermost region of the accretion flow is correct, GRAVITY has the potential of directly determining the space-time metric around this black hole. GRAVITY thus may be able test General Relativity in the presently unexplored strong curvature limit close to the event horizon [6].

- GRAVITY will be able to unambiguously detect and measure the mass of black holes in Globular clusters throughout the Milky Way (such as ω Cen [8]) by means of stellar dynamics.

- For those AGN for which the sphere of influence of the SMBH is resolved, GRAVITY will enable direct measurements of the mass of the SMBH by determining the dynamics of the Broad Line Region (BLR, [9]).

- Measuring a velocity gradient in the BLR also allows to determine the size of the BLR, and thus enables one to test the scaling relation linking the luminosity of the AGN to the size of the BLR [10].

- GRAVITY can be used to test nuclear star formation in AGN on scales a factor of 10 smaller than what is possible nowadays with AO [11].

- For Young Stellar Objects (YSO) resolved spectroscopy allows to probe the size and dynamics of the hydrogen Bracket gamma emission from the gas in the accretion disk across the whole initial mass function [12].

- Another ambitious goal for GRAVITY are jet morphology studies in YSO [13]. GRAVITY will be able to resolve the central jet formation engine around young, nearby stars. Achieving 10 μas in one hour gives access to motions in the jets of T Tauri stars, moving at ~ 150 km/s.

- GRAVITY will be able to probe the nature of any extended NIR emission for the sample of known microquasars (with reference stars) in fine detail [14].

- Furthermore, GRAVITY will explore classical, interferometric objects, such as binary stars, exoplanet systems and young stellar disks.

GRAVITY thus will carry out a number of fundamental experiments, as well as increase substantially the range and number of astronomical objects that can be studied with the VLTI. Figure 2 illustrates what astrometry at the 10 μas level means in terms proper motion and acceleration measurements.

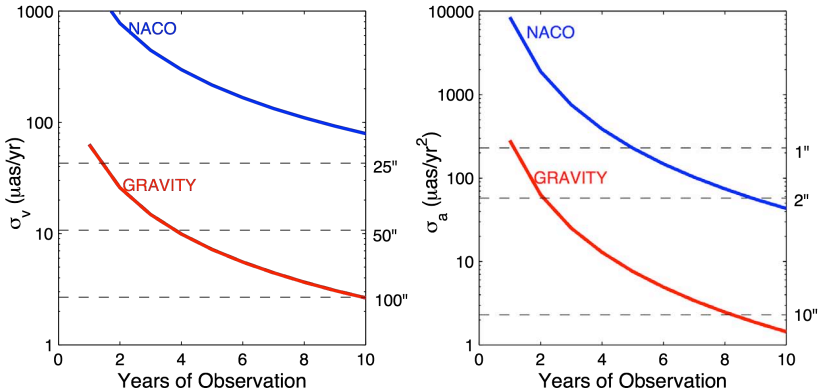

Figure 2: Proper motion (left) and acceleration (right) accuracies (5σ) that will be achieved with GRAVITY, as a function of the length of the observing period, and compared to the accuracies that can be obtained with NACO/VLT AO imaging. An intrinsic 1σ astrometric uncertainty of 10 μas is assumed, with 3 − 4 measurement periods per year. In each inset, the label on the right vertical axis denotes the largest distance R from Sgr A* to which such proper motions and accelerations are expected.

## 2.2 Comparison with other facilities

The VLTI is the only array of its class offering a large 2" field of view (40 times the diffraction limit of the 8 m Unit Telescopes in K-band). GRAVITY will for the first time utilize this unique 2" field of view, providing simultaneous interferometry of two objects for four telescopes. This allows narrow angle astrometry with a precision of 10 μas, using six baselines at a time. The combination of large apertures and six baselines is unparalleled by any other interferometer.

A second new and unique element of GRAVITY is the use of infrared wavefront sensors to open a new window for interferometry. None of the competitive interferometric arrays has presently planned or will be able to upgrade their adaptive optics for infrared wavefront sensing.

The application of phase-referenced imaging – instead of closure phases – is a major advantage in terms of model-independence and fiducial quality of interferometric maps with a sparse array such as the VLTI.

## 2.3 The physics perspective of testing general relativity

It is well worth to compare the route of testing general relativity with GRAVITY with other experiments.

- Classical solar system tests give access to the low curvature, low mass regime of general relativity [6].
- Earth-bound gravitational wave detectors such as LIGO, VIRGO or GEO should be able to detect single supernova explosions, corresponding to the high-curvature, low-mass regime of GR. The high-mass regime requires going to lower frequencies, which are not accessible from ground.
- Space-borne gravitational wave detectors (e.g. LISA) will extend the accessible frequency range to the regime in which the signal of merging MBHs is expected to occur, testing the high-curvature, high-mass regime of GR.
- A submm-VLBI array should be able to actually resolve Sgr A*, possibly showing its event horizon as a shadow [15].

The so far untested strong-field limit of general relativity is probed with the last three items only. Gravitational wave detectors have a different focus, since very dynamic events will be observed. Thus the structure of space-time is tested rather indirectly. Furthermore gravitational wave detectors are quite expansive devices, and only upper limits on the emission of gravitational waves have been obtained so far experimentally. The submm-VLBI observations are unlikely to yield a dynamic picture of Sgr A* since the estimates for the exposure times needed are much longer than the orbital period, the characteristic time scale of the system. Hence, GRAVITY offers a quite promising route to severely test general relativity. It can be considered as a test particle approach to the structure of space-time around a MBH, which at the same time has only a very moderate price, namely that of building a 4-telescope interferometric beam combiner.

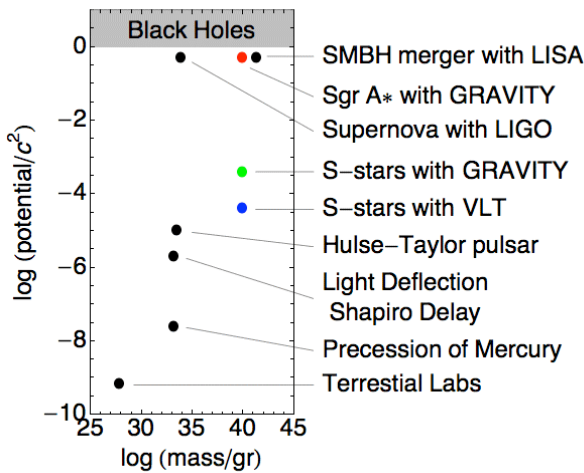

Figure 3: Dynamical tests of general relativity as a function of the mass of the gravitating body and the curvature of space-time. Adapted from [6].

# 3. KEY SCIENCE CASES

## 3.1 Stellar orbits around Sgr A*

The combination of adaptive optics (AO) and large (8 m – 10 m) telescopes made it possible to observe a multitude of stellar orbits moving in the gravitational potential of the MBH Sgr A* in the Galactic Center. Up to now the system can be described perfectly by a single point mass and Newtonian gravity [5]. Nevertheless, deviations from these simple assumptions are expected to exist: A cluster of dark objects with stellar masses (e.g. neutron stars or stellar mass black holes) might well be present around Sgr A* [16] yielding deviations from the single point mass hypothesis. Furthermore, the effects of general relativity will break the assumption of a Newtonian system. In order to detect such deviations, NIR interferometry is a suitable tool.

We have examined the feasibility of detecting the Schwarzschild precession around Sgr A*. Given the density profile and luminosity function of the Galactic Center star cluster [17], we estimate that a few (three to eight) stars with $17 < m_K < 19$ at any point in time should reside in the central 100 mas, thus being essentially unresolved with current NIR instrumentation, but being accessible with the VLTI. Simulating observations using all four UTs for 9 hours, we were able to show that it is possible to recover the assumed star fields from the simulated data. Such stars will have orbital periods of ~ 1 year and their orbits should precess by a few degrees per revolution.

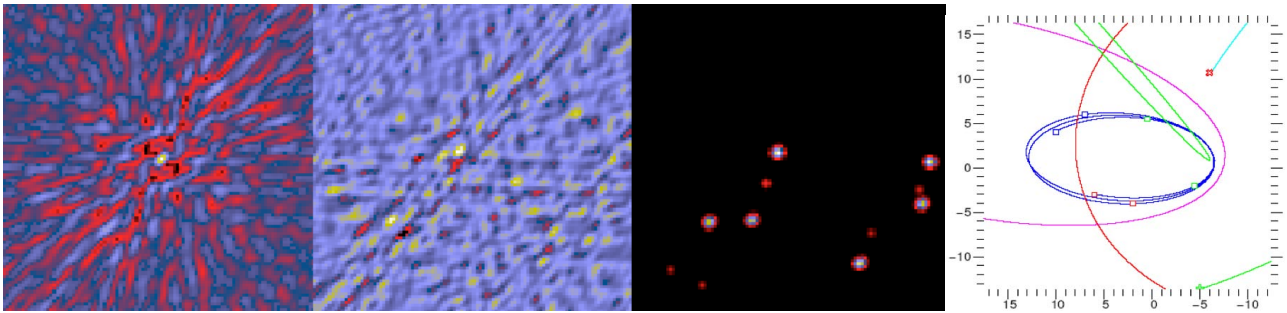

Figure 4: Simulated observation of a star field with 6 stars placed in orbit around Sgr A* in a 100 mas square field. Left: The PSF for one night of VLTI observations. Middle-left: The reconstructed image. Middle-right: The recovered image from the middle panel, using a simple CLEAN algorithm. Right: Simulated orbit figures (in mas) for the stars using images from several epochs. The strong precession due to the Schwarzschild metric is evident after even only two revolutions, each lasting no more than a year.

We have investigated the precision by which stellar positions can be retrieved from the reconstructed images. The astrometric precision thus obtained is of the order of the Schwarzschild radius for all sources in the field of view. These imaging based results can be used to bootstrap a fit on the complex visibilities. They allow fixing the number of stars and provide an initial guess for their position and brightness. Such a fit yields precisions below 1 µas for an $m_K = 12$ source, or 5 µas for an $m_K = 17$ source. At this level, unaccounted for systematic effects will be predominating. The astrometric precision is very dependent on the number of sources in the field. The statistical properties of the stars at the centre of the Galaxy make it plausible that three stars could be observable at any time in GRAVITY's field of view, which still yields a precision of 10 µas.

## 3.2 Flares from Sgr A*

Genzel et al. 2003 [18] observed for the first time sporadic NIR emission from Sgr A* (figure 5). Since then, many such flares occurring at a rate of 1/night and lasting each for ~ 2 hours have been observed. Most of the flares show a quasiperiodic substructure with a typical time scale of ~20 minutes [19] This can be understood in an orbiting hot spot model, where a heated gas blob close to the innermost circular orbit revolves around the MBH, yielding the light curve modulations due to the orbiting motion (figure 5). The emission arises probably due to a magnetic reconnection event in the accretion flow, during which the energy from the magnetic field is heating electrons locally [20]. The size of the flare region is actually very small, given the typical rise times of few minutes only.

Since the apparent diameter for the event horizon a MBH of 4 million solar masses at a distance of 8 kpc is 10 µas, the motion of a bright, compact source at the last stable orbit with a diameter of 60 µas might be detectable with NIR interferometric means. The flares will not be resolved, however the astrometric wobble of the centroid can be detected.

While such an observation would already be extremely exciting, an even more rewarding goal would be to characterize the motion. The spin of the MBH and strong lensing effects will lead to characteristic deviations from a circular motion. Thus, the flares are used then as probes for the strongly curved space-time in which they move, ultimately testing the theory of gravity, general relativity, in its strong field limit and for an extremely heavy mass. This perspective coined the name GRAVITY for the instrument.

We have simulated VLTI observations of flares from Sgr A*. Assuming reasonable parameters for the beam-combining instrument, we were able to show that already the observation of a single flare will allow us to detect the orbital motion. If ~ 10 flares can be coadded suitably, the strong relativistic effects become visible [21, 22]. It should also be noted that neither the exact emission mechanism nor the exact motion are strong prerequisites. Given that flares occur close to the event horizon, their velocity should be of order speed of light, resulting in a total traveled path for a flare of one hour of ~500 µas. This number is sufficiently larger than the envisioned accuracy of 10 µas, showing the potential for GRAVITY to actually map the motion with high precision.

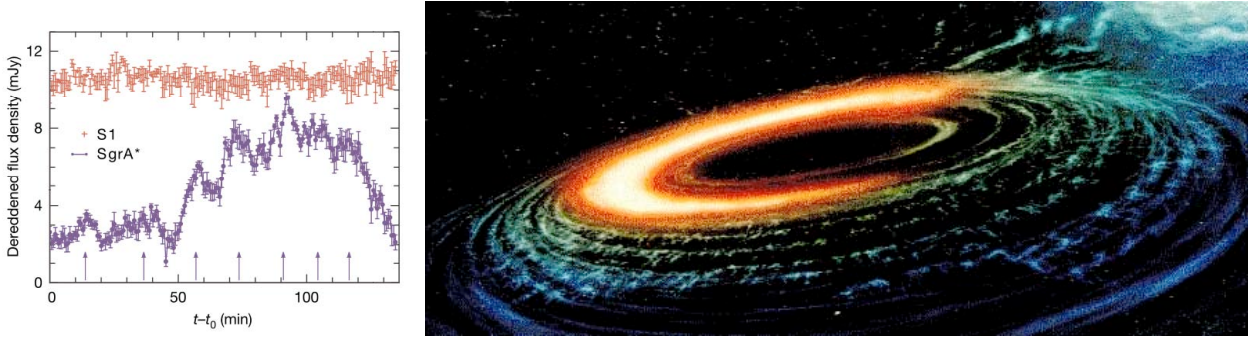

Figure 5: Left: Light curve of the first NIR flare from Sgr A*, showing a characteristic substructure, denoted by the arrows at the x-axis. Right: Artist's impression of a flare orbiting at the last stable orbit around the SMBH.

## 3.3 Active Galactic Nuclei

The standard unified model [23] postulates that all active galactic nuclei (AGN) are accreting SMBHs surrounded by a geometrically thick, dusty interstellar cloud structure (the 'torus') whose orientation relative to the observer's line of sight determines the specific phenomena observed. Historically it has proven difficult to study AGN at the spatial scales on which these components exist. For seeing limited observations, at a distance of 20 Mpc, 1" corresponds to 100 pc. Using AO for NIR observations, it has become possible to probe these objects at 100 mas scales. MIR interferometry has now resolved the dusty torus for roughly ten AGN and even more detailed views were possible for NGC 1068 [24, 25] and NGC 4151 [26]. GRAVITY will continue that route to smaller angular sizes. In the closest AGN – those within 20 Mpc – one will be able to probe scales less than 0.5 pc. And in Circinus and Cen A, the spatial resolution will be around 0.1 pc, a scale which is close to what was possible in the Galactic Center before the advent of adaptive optics.

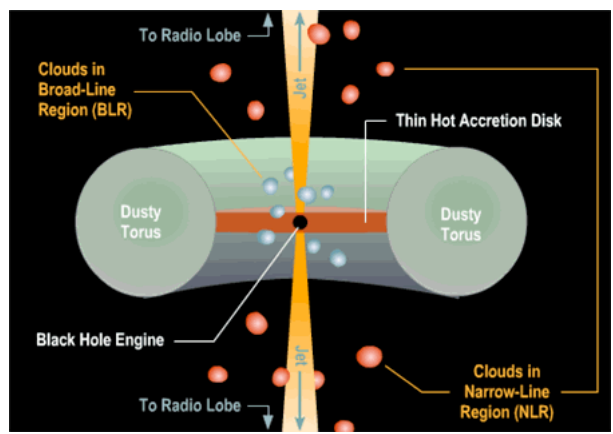

Figure 6: Schematic view of the unified AGN model

Perhaps the most exciting advances for AGN science that will become possible with GRAVITY concern the Broad Line Region (BLR). This is a compact region lying between the AGN and the inner edge of the torus, in which the large width of optical/NIR emission lines arises from the high velocities of the clouds as they orbit the central BH. Currently, the size of the BLR can only be inferred indirectly from time variability studies ('reverberation mapping', available for 35 AGN [27]), where BLR sizes are derived from the time delay between UV continuum and emission line variations. Despite the BLR being always < 0.1 mas across and hence too small to resolve with VLTI direct broad-band imaging at infrared wavelengths, the excellent 10 µas astrometric accuracy of GRAVITY will enable deriving a velocity gradient across it. First and foremost, this will provide a statistical estimate of the fraction of BLRs in which there is a significant component of ordered rotation. And then, for these sources, it will become possible to make an estimate of the size of the BLR and also directly determine the central BH mass.

Another important aspect is nuclear star formation. There are observational and theoretical reasons to believe that the AGN and the surrounding star formation are influencing each other. This interaction can have an impact on different scales, from fuelling the AGN (because of the effect a nuclear starburst will have on gas inflow) to the evolution of the galaxy (via feedback from the AGN). GRAVITY's spectroscopic capabilities will allow studying this on the relevant physical size scales.

### 3.4 Intermediate mass black holes

There is general consensus that massive black holes are ubiquitous in galactic nuclei down to masses of a few $10^5$ $M_\odot$, and there appears to be a fairly tight correlation of the mass of the central BH with the mass (or velocity dispersion) of the spheroidal stellar component associated with the galaxy (bulge or elliptical, [28]). This correlation suggests that during the violent/rapid formation of a spheroidal stellar system about 0.1–1% of the mass collects in the centre in form of a black hole. Theoretical simulations imply that the same process may also lead to the formation of intermediate mass black holes (IMBHs: a few $10^2$ to a few $10^4$ $M_\odot$) in sufficiently massive and dense star clusters [29], as the result of core collapse and collisional build-up of a central object. Such IMBHs may also be of relevance as "seeds" of the SMBH population, perhaps formed at redshifts z > 10 in dense population III star clusters.

Recent searches in globular clusters have provided tantalizing but contradictory evidence for such IMBHs [30, 31, 32]. The main reason for this unclear situation is the fact that the sphere of influence of the postulated BHs is less than a few arcseconds typically, such that only a few stars are available for a statistical determination (e.g. of the radial velocity dispersion), naturally leading to large Poisson errors.

GRAVITY may dramatically change this situation in a few suitable cases. With its high angular resolution, proper motions of faint stars very close to the centre may be obtained and most importantly, accelerations may be determined. Under certain simplifying assumptions even a single acceleration measurement can exclude/prove the existence of a high mass within the inferred orbit of the star. Determination of several accelerations can be used as a precise tool for determining the gravitational potential, as in the Galactic Center.

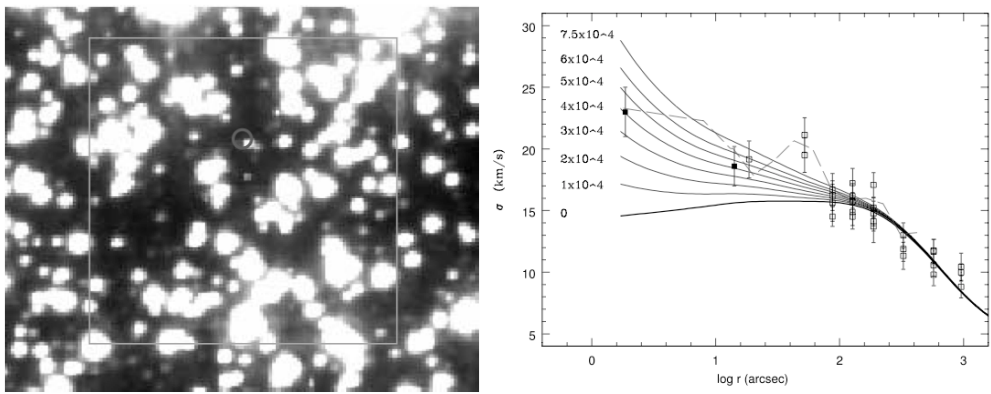

Figure 7: Left: The central region of the globular cluster ω Cen, observed with ACS on board the Hubble Space Telescope. Right: Radial velocity profile, showing a characteristic increase towards the center, compatible with an IMBH of a mass of 40,000 $M_\odot$ [8].

## 3.5 Young Stellar Objects

Protoplanetary disks around young stellar objects (YSO) are complex systems of many interacting components. These young accreting systems are composed of a central protostar or pre-main sequence star surrounded by a disk of gas and dust, and some are known to drive powerful jets and outflows. The star is magnetically active with powerful X-ray emission from the stellar magnetosphere, including large flares. The stellar magnetosphere may interact with the disk. Accretion onto the star is observed, and is commonly accepted that it is magnetically channeled Complex chemistry takes place in the disk, from the irradiated disk surface to the deep disk interior.

GRAVITY will be ideal for investigations of YSOs, namely their circumstellar disks and outflows. Since the systems are embedded, GRAVITY's infrared wavefront-sensing capability is critical. The disk may for example show spiral structures, wakes and gaps expected as a result of the interaction between a forming planet and the disk. Such structures have indeed already been observed in a number of cases (e.g. GG Tau, Fomalhaut). GRAVITY's 4 mas resolution (compared to 60 mas resolution for one UT) drastically improves the contrast between a disk and its embedded planet. It should be possible to probe for young giant planets, which are almost 6 mag fainter than what is currently achievable.

The physics behind the formation of jets in YSOs, and in particular the launching mechanism, is still poorly understood. The important processes seem to take place within less than 0.5 AU from the star, which at typical distances to the nearest star forming regions of 150 pc translates into an angular size of less than 30 mas. At 4 mas resolution, GRAVITY will be able to resolve the central jet formation engine around young, nearby stars. Furthermore, at a distance of 150 pc, an astrometric precision of 10 µas over a time span of 1 h corresponds to a transversal velocity accuracy of $\approx$ 60 km/s. Hence, high-velocity outflows and the formation and evolution of jets from T Tauri stars with typical velocities of 150 km/s can be resolved and the time evolution of jet formation at the base of the outflow can be directly traced with GRAVITY [13].

Using the spectral resolution of R ~ 5000 GRAVITY will be able to probe the size and dynamics of the hydrogen Bracket gamma emission across the initial mass function spectrum up to the hydrogen-burning limit. From this, accretion rates can be measured from the equivalent width of the line. The resulting geometry and dynamics can then be directly compared to radiative transfer simulations of jet engine models.

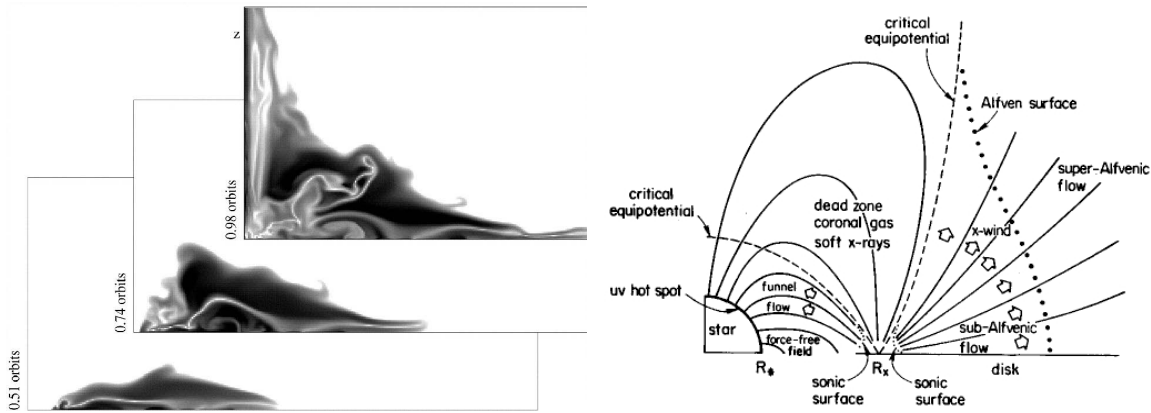

Figure 8: Left: logarithmic density structure for a recently launched outflow driven by an internal magnetic disk field [13]. Each box is 0.5 × 0.25 AU$^2$, corresponding to 10 × 5 mas$^2$ at 50 pc. GRAVITY would aim at resolving these structures, and tracing their time evolution over several weeks to months. Right: schematic of the disk-outflow structure in young stellar objects [33].

## 4. TOP LEVEL REQUIREMENTS

Given the key science cases we have derived the following top-level requirements:

- Operation in the K-band (2.2 µm), offering a well-suited atmospheric window and being the scientifically most interesting waveband for all types of obscured objects.
- Interferometric combination of the light collected by the four UTs of the VLT; mainly driven by the desired UV-coverage and faintness of the sources.
- Simultaneous combination of the light of two sources for each of the six baselines.
- Adaptive Optics (AO) at each telescope using a wavefront sensor operating in the NIR. A Strehl ratio of ~35% should be reached for a magnitude of $m_K = 6.5$ assuming typical atmospheric conditions for the VLTI site, Paranal, and a distance of the AO guide star to the science field of 6". Wavefront sensing should be possible also on-source.
- Beam stabilization from the deformable mirror of the AO to the beam combiner instrument.
- Fringe Tracking, either on source or using a phase reference source within the field of view of the VLTI, down to $m_K = 10$ assuming typical atmospheric conditions.
- Control of the relevant optical path differences at the level of 5 nm, corresponding to the desired accuracy of 10 µas.
- Three spectral resolutions: R = 22, 500, 4000, corresponding to 5, 128, 1024 resolution elements over the K-band.
- Active polarization control.

## 5. WORKING PRINCIPLE

The working principle of GRAVITY is explained for the case of Galactic Center observations (figure 9). The field offers a bright AO reference star (IRS 7, 5.5" separation from Sgr A*, $m_K = 6.5$) outside the 2" field of view of the VLTI, and a fringe-tracking star (IRS 16C, 1.2" separation from Sgr A*, $m_K = 9.7$) inside the field of view.

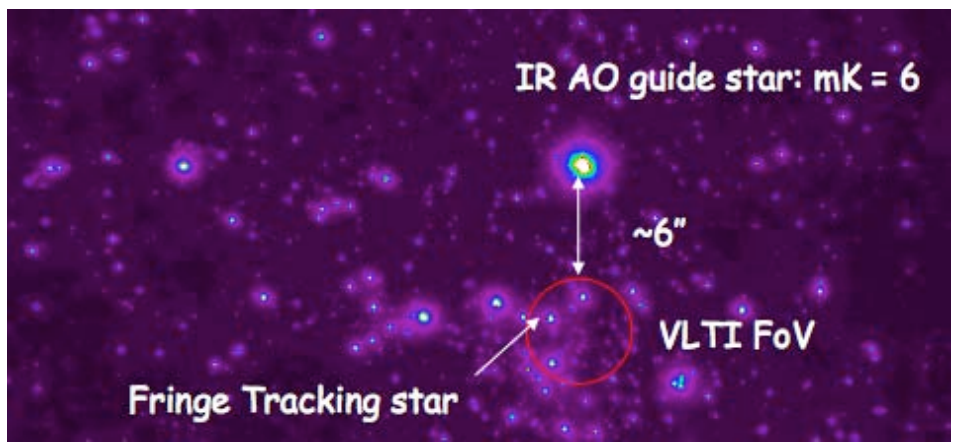

Figure 9: The Galactic Center field

The following description refers to figure 10, where the working principle is illustrated for one of the six baselines.

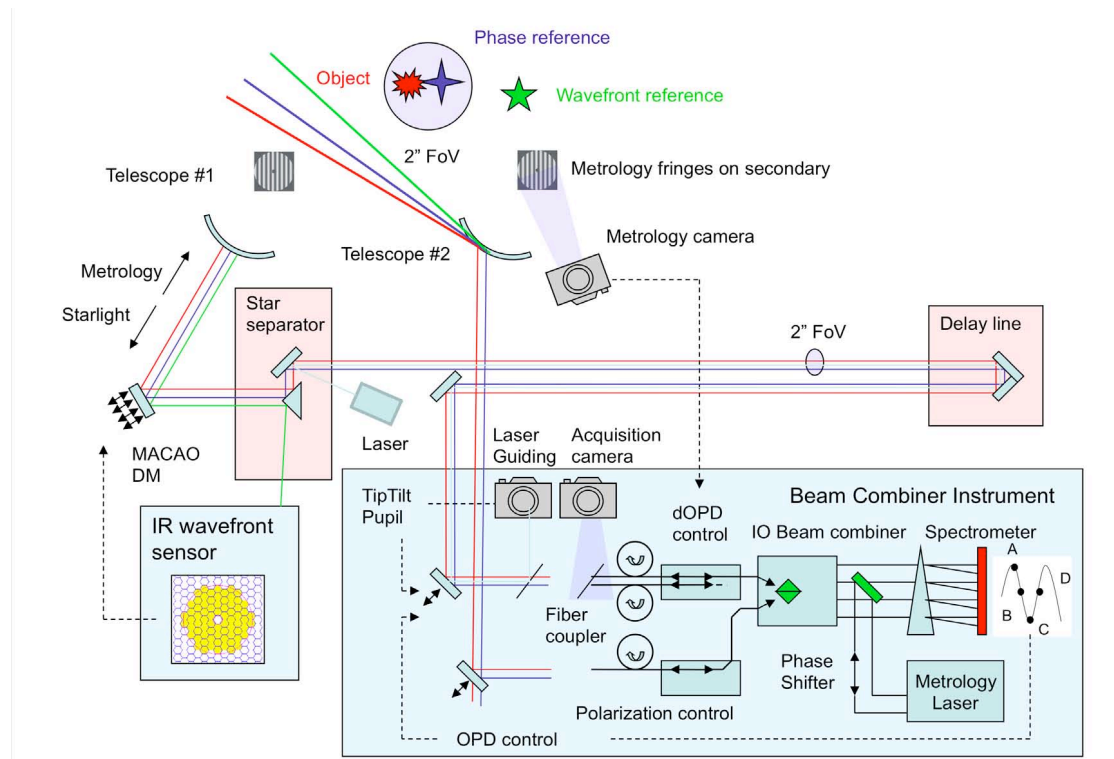

Figure 10: Working principle of GRAVITY

GRAVITY provides four (J+)H+K-band (tbd) infrared wavefront sensors, located at the Coude rooms of the UTs. The AO reference star is picked with the PRIMA star separator and imaged onto the GRAVITY IR wavefront sensors. Their signals will be used to command the existing MACAO deformable mirrors. The system will compensate atmospheric turbulence and can work on either of the two beams (on-axis or off-axis) after the star separators.

Behind the adaptive optics, the VLTI induces additional high-frequency tip/tilt and pupil errors. These are in turn corrected by a dedicated laser guiding system. The laser beams are launched at the star separators and trace the optical path down to the beam combiner instrument in the VLTI lab. Furthermore, low frequency drifts of the field and pupil can occur (for example due to the delay line motion). These will be measured and corrected at a low bandwidth with the camera working at H-band that is used during the acquisition process. The interplay of all these systems serves solely the purpose to bring an unperturbed and seeing-corrected beam to the VLTI lab.

The 2" field of view of the VLTI contains both the science target (Sgr A*) and the phase-reference star (IRS16C). Both objects are re-imaged via the main delay lines to the GRAVITY beam combiner instrument, a block diagram of which is shown in figure 11. The fiber coupler then splits the light of the two stars and injects it into single mode fibers. A rotatable half-wave plate is used to analyze the linear polarization of the light. A fiber control unit including rotators and stretchers align the polarization for maximum contrast, and compensate the differential OPD between the phase reference star and science object – caused by the angular separation on sky. The beam combiner itself is implemented in an integrated optics chip with instantaneous fringe sampling.

The bright reference star (IRS 16C) feeds the fringe tracker. The OPD correction is applied to an internal small internal piezo-driven mirror stabilizing the fringes of both the reference star and the faint science object (Sgr A*) at high frequency. Low-frequency corrections will be offloaded to the main delay line. The science spectrometer is optimized for longer, background limited integration times of faint objects, and offers a variety of operation modes, including low to moderate (R ~ 4000) resolution spectroscopy and polarization splitting.

The differential OPD between the science and reference beam is measured with a laser metrology system. The laser light is back-propagated from the GRAVITY beam combiners covering the full beam up to the telescope secondary mirror, producing there a fringe pattern, which carries the dOPD information. This fringe pattern is observed in scattered light using an infrared camera mounted to the UT telescopes, or in direct light using photo-diodes at the AT telescopes. Together with the Metrology signal, GRAVITY provides simultaneously for each spectral channel the visibility of the reference and science object, and the differential phase between reference and science object. The GRAVITY data can be used for interferometric imaging exploring visibilities and closure phases from six baselines, and for astrometry using the differential phase- and group delay.

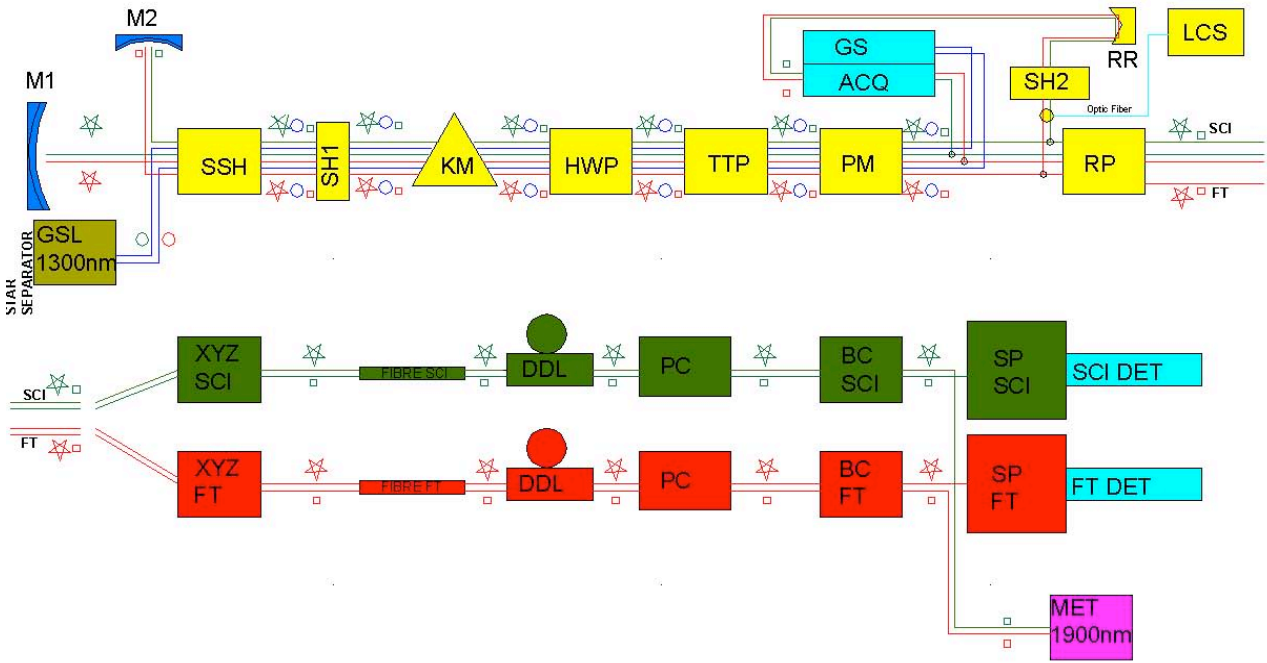

Figure 11: Block diagram of the GRAVITY beam combiner instrument: GSL: guiding system laser, SSH: safety shutter, SH1: entrance shutter, KM: K-mirror, HWP: half wave plate, TTP: tip tilt piston mirror, PM: pupil actuator mirror, GS: guiding system sensor, ACQ: acquisition camera, RR: retro-reflector, SH2: shutter for retro-reflector, LCS: metrology laser diode, RP: roof prism, XYZ SCI: x,y,z stage of science fiber, XYZ FT: x,y,z stage of fringe tracker fiber, FIBRE SCI: science fiber, FIBRE FT: fringe tracking fiber, DDL: differential delay line, PC: polarization controller, BC SCI: science beam combiner, BC FT: fringe tracking beam combiner, SP SCI: science spectrometer, SP FT: fringe tracking spectrometer, SCI DET: science detector, FT DET: fringe tracking detector, MET: metrology laser

## 6. PRELIMINARY DESIGN

### 6.1 General Layout

Figure 12 (left) shows the location of the GRAVITY beam combiner instrument, feeding optics and electronics cabinets in the VLTI lab. All functions of the GRAVITY beam combiner instrument are implemented in a single cryostat for optimum stability, cleanliness, and thermal background suppression (figure 12, right). The cryostat hosts all functions of the beam combiner instrument. It provides the required temperature for the various subunits with temperatures ranging from 80 K up to 290 K. The bath-cryostat is cooled with liquid nitrogen, and makes use of the gaseous exhaust to cool the intermediate temperature (240 K) subsystems. All temperatures levels are actively stabilized with electric heaters. The cold bench is supported separately from the vacuum vessel and liquid nitrogen reservoir to minimize vibrations within the instrument.

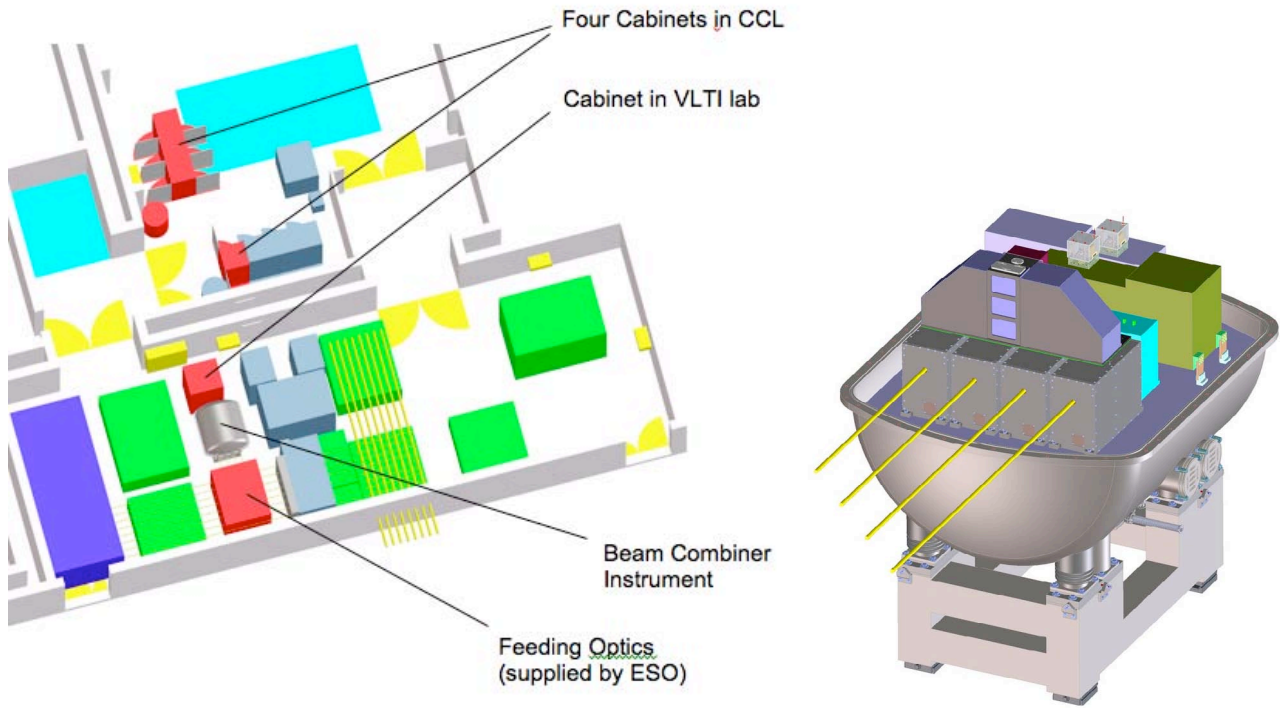

Figure 12: Left: The planned layout of the VLTI laboratory at the time when GRAVITY will be installed. Right: Preliminary design of the cryogenic vessel.

## 6.2 Adaptive Optics

The wavefront sensor of GRAVITY consists of a warm part, the adaptive optics mode selector, and a cold part including the main components of a Shack-Hartmann wavefront sensor, in particular the lenslet mask and the detector. The locations of the GRAVITY wavefront sensors will be at the Coudé focal station of each unit telescope, in the same room as the already existing MACAO systems.

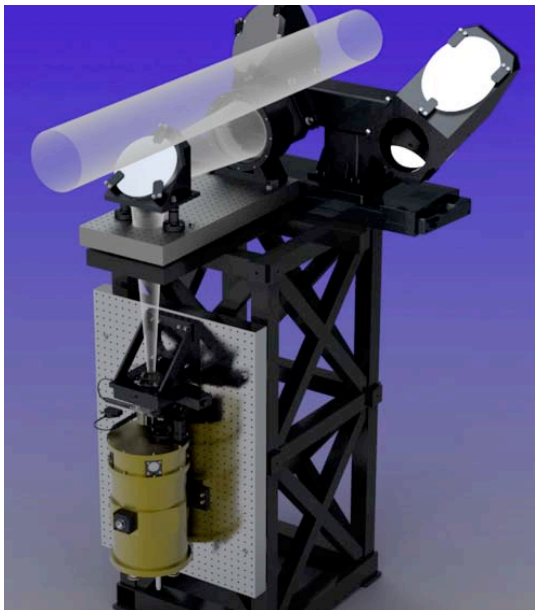

| AO Reference Star Data | Expected performance | Requirement |
|---|---|---|
| Adaptive optics on K=7 star (on-axis, 30° zenith angle) | 45 % Strehl | 35 % Strehl |
| Adaptive optics on K=7 star (7" off axis, 30° zenith angle) | 33% - 38% Strehl | 25 % Strehl |
| Adaptive optics on K=10 star (on-axis at zenith) | 13 % Strehl | 10% Strehl |

Table 1: Expected performance of the GRAVITY adaptive optics system at the Coudé focus. The values applicable for the beam combiner instrument in the VLTI lab will suffer additionally from the uncorrected aberrations and the residual tunnel seeing.

Figure 13: Left: Opto-mechanical design of the GRAVITY wavefront sensor. The detector and lenslet array is housed in the small cryostat to the bottom left.

The wavefront sensors will be of Shack-Hartmann-type, with 9 × 9 subapertures, using (J+)H+K-band (tbd) light. The system is similar to NAOS [34]. The correcting element of the adaptive optics will be the existing deformable mirrors of the MACAO systems, which currently are used for adaptive optics in the visible. We have simulated the expected performance (table 1), showing that the design fulfills the requirements [35]. The limits are set by the detector readnoise. The limiting magnitude can be improved by ~ 0.75 mag with the next generation detectors presently under development.

### 6.3 Beam Combiner Pre-Optics

The fiber coupler (figure 14, [36]) feeds the light from the reference and science objects into the fibers. A motorized K-mirror corrects the rotation induced by the VLTI optical train. A motorized half-wave plate allows the independent rotation of the linear polarization. An off-axis parabolic mirror relay optics focuses the star-light onto a roof-prism to separate the reference and science star. Two separate relay optics then couple the reference and science starlight into their respective fibers. The fibers are mounted on piezo-driven x,y,z-stages to pick the objects and adjust the focus. A piezo driven θ,ϕ,z-mirror provides tip-tilt and piston control. Another piezo-driven mirror provides lateral pupil control. A dichroic beam splitter separates the light to feed the fibers ($> 1.9\ \mu m$) and the acquisition and guiding cameras ($< 1.9\ \mu m$). A cat-eye reflector behind the dichroic allows imaging the fiber entrance (launching the laser metrology) onto the acquisition camera (section 6.7). Two motorized shutters pick either the starlight or the metrology light for the acquisition camera. A photodiode behind the dichroic provides an internal reference signal for the laser metrology.

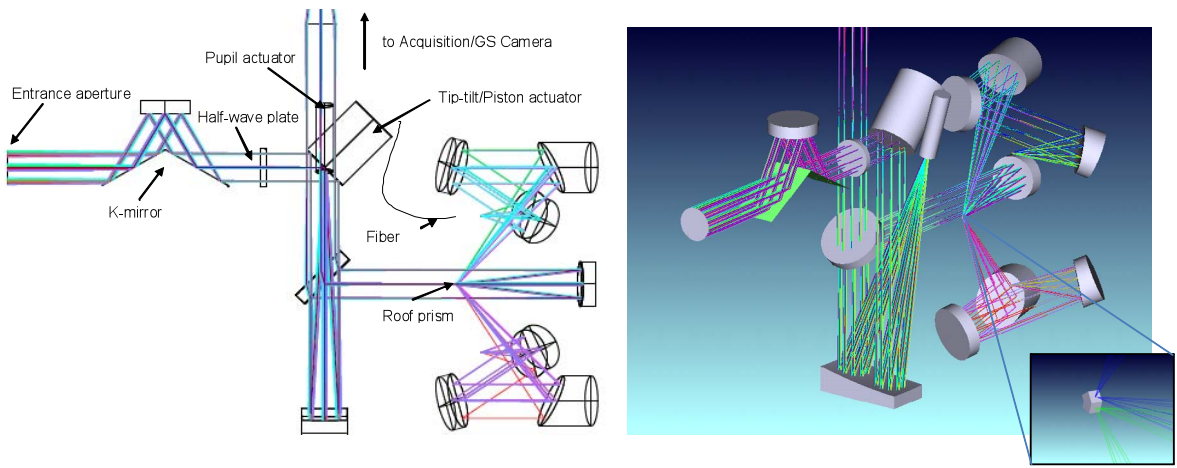

Figure 14: Optical Layout of the fiber coupler unit.

### 6.4 Fiber Optics

The optical fibers provide single-mode filtering of the incoming wavefront and transport the light to the integrated optics beam combiner. GRAVITY uses non-birefringent single-mode Fluoride fibers with negligible absorption in the K-band. The fiber control unit provides the functions to adjust the polarization rotation angle and to adjust the optical path length. The random rotation of the polarization by the fibers is compensated with motorized polarization rotators to align the polarization of each telescope for optimum fringe contrast (figure 15, left). The optical path length of the fibers can be adjusted by stretching the fibers, which are wrapped around a piezo ceramic cylinder (figure 15, right).

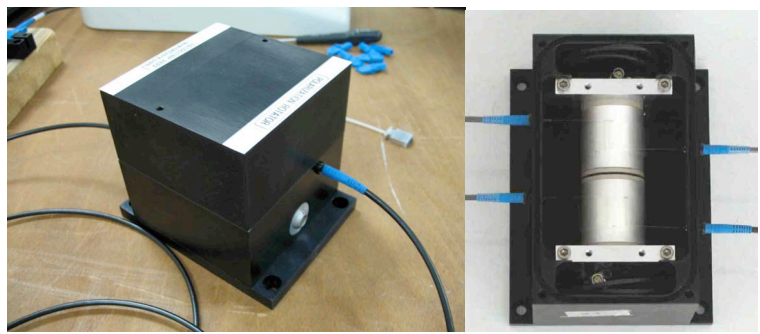

Figure 15: Prototype polarization rotator (left) and fiber-differential delay line (right)

## 6.5 Beam Combiner

The two beam combiners for the reference star and the science object are implemented in integrated optics. The beam-combiner chips are directly fed by the single mode fibers, and provide instantaneous pair-wise combination of all six baselines for the four telescopes. Internal phase-shifter and splitter provide the instantaneous ABCD sampling of each interferogram. As such the integrated optics beam combiner provide 6 (baselines) × 4 (samples) = 24 outputs for the four inputs (figure 16, [37]).

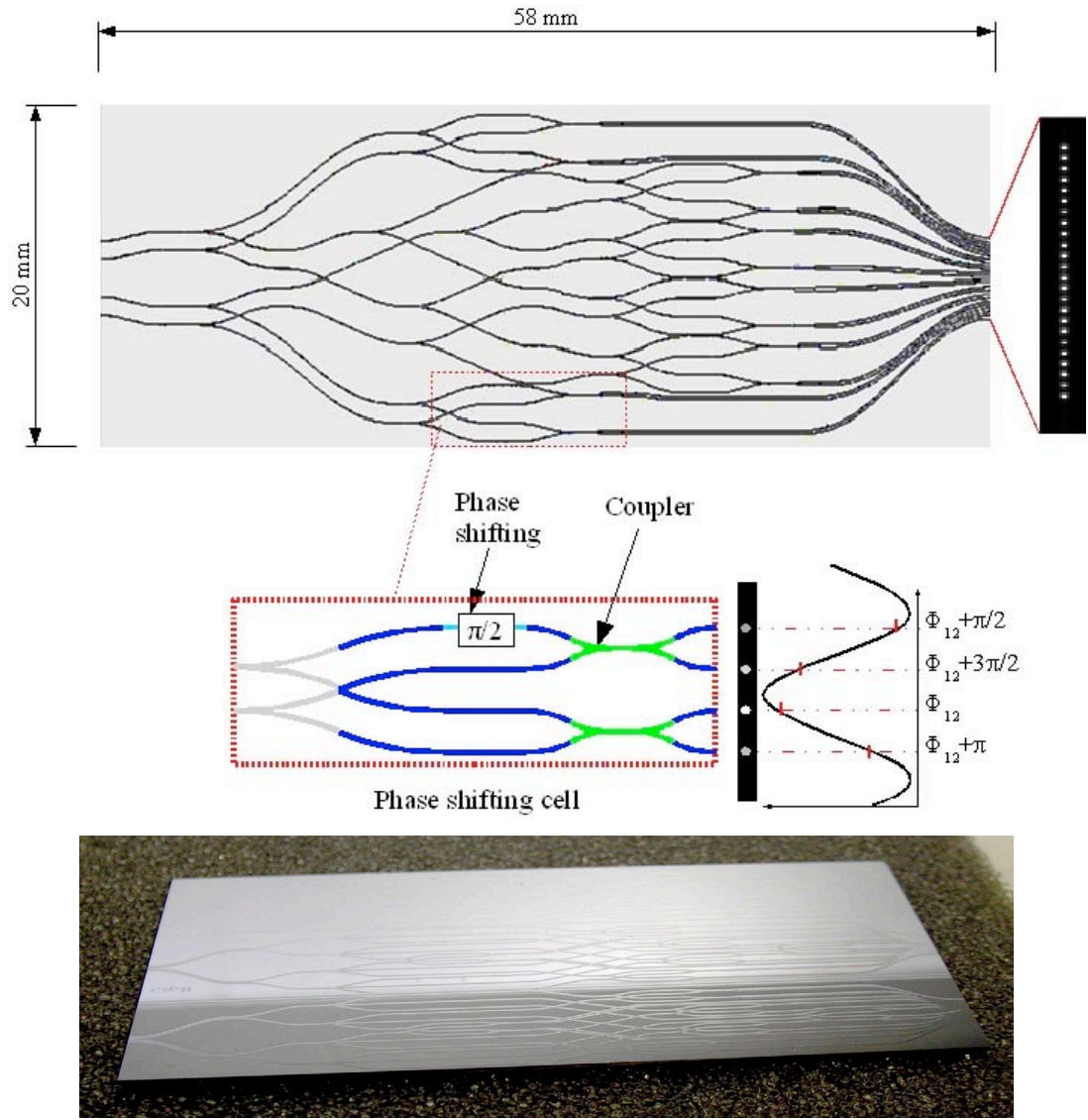

Figure 16: Top: Beam combination principle for the 4-telescope, 6-baseline integrated optics beam combiner. Bottom: Prototype beam combiner, available in the lab.

## 6.6 Spectrometers

Each of the integrated optic chips feeds a spectrometer. One spectrometer is optimized for high speed and throughput to serve the fringe-tracker using R = 22, the other spectrometer provides additional high spectral resolution (R = 500, 4000) and serves long exposures. Both spectrometer follow the classical collimator – dispersing element – camera concept, and are implemented in a lens design [38, 39, 40]. The dispersing elements are glass prisms for the R = 22 spectra, the higher spectral resolutions are achieved with grisms. Both spectrometers are equipped with Wollaston prisms to split linear polarization. The spectrometers also provide the optics to couple the metrology laser into the integrated optics, and ultra-high optical density filters to suppress backscattering of the laser metrology onto the detectors (figure 17, [41]). The science spectrometer detector is a ~ 2048 × 512 area of an HAWAII2RG detector, operated with 32 output channels and a sampling rate of 100 kHz. The fringe tracker detector is planned to be a SELEX APD detector with 24 µm pixels, using ~ 128 × 40 pixels. The fringe tracker runs nominally at 350 Hz frame rate, but should be able to go as fast as 800 Hz.

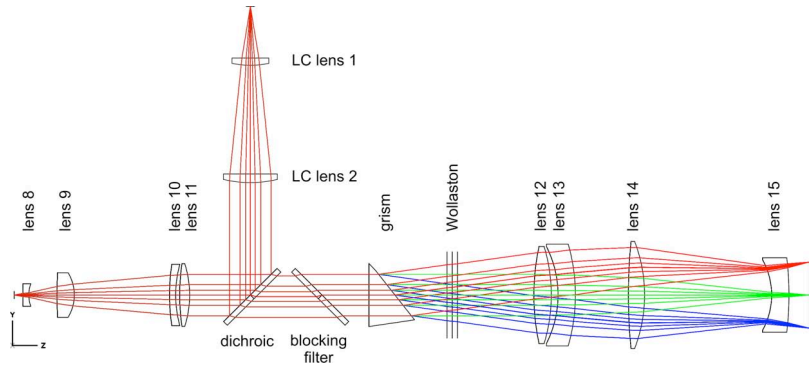

Figure 17: Optical layout of the spectrometers.

## 6.7 Acquisition Camera

The acquisition camera provides simultaneously an image of the focal plane, a pupil image, and a Shack-Hartmann wavefront sensor image for all four telescopes. It is used to acquire the object, to analyze the beam quality, and to track low-frequency drifts of the field and pupil. The acquisition camera is fed by a dichroic beam-splitter within the fiber coupler optics, which is located after the tip/tilt/piston and pupil-actuator to allow a closed loop control (figure 18, [42]). The acquisition camera detector is a ~ 768 × 1024 area of an HAWAII2RG detector, operated in the 32 output channel mode, i.e. using 16 output channels, with a sampling rate of 100 kHz.

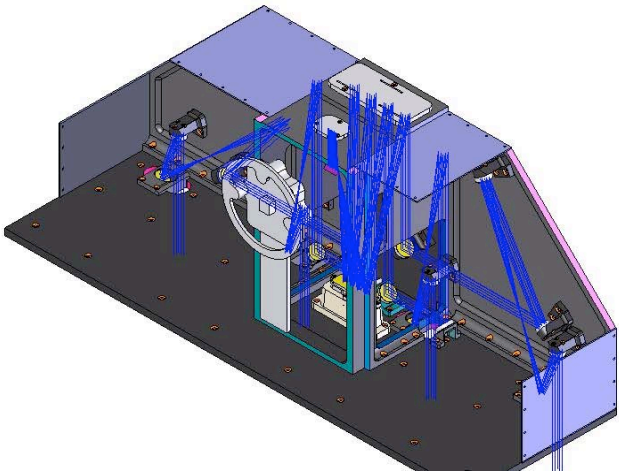

Figure 18: Optomechanical layout of the acquisition camera

## 6.8 Guiding System

The guiding system corrects for high-frequency tip/tilt and lateral pupil motion introduced in the VLTI optical train between the star-separators and the GRAVITY internal fiber coupler. This is done by launching two laser beacons – one in the pupil, one in the field – at each star-separator, and sensing the beam-motion with position-sensitive-diodes in the beam combiner instrument. To avoid additional optical elements in the optical train of the star-separators, the laser-beams are generated via scattering of laser beams focused on the pupil and field mirrors, respectively (figure 19, [42])

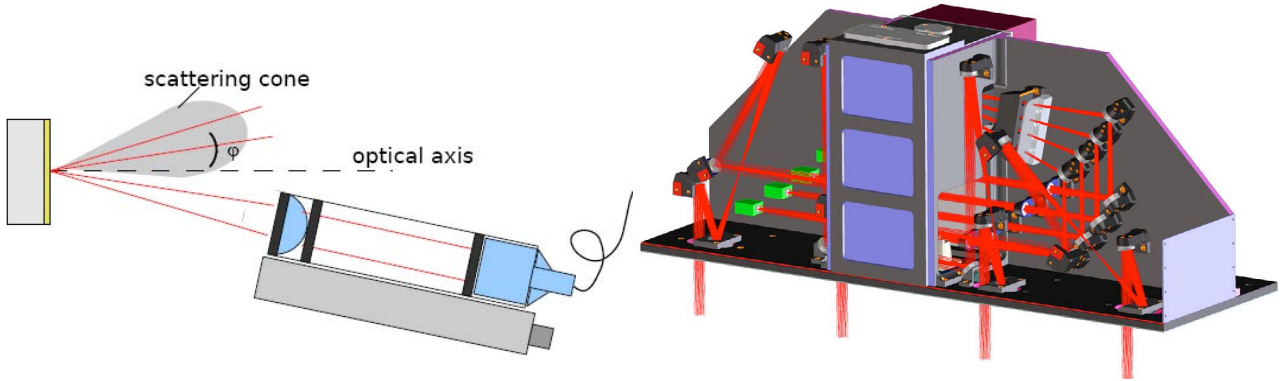

Figure 19: Left: Illustration of the laser launching principle. Right: Opto-mechanical design of the guiding system sensor inside the beam combiner instrument.

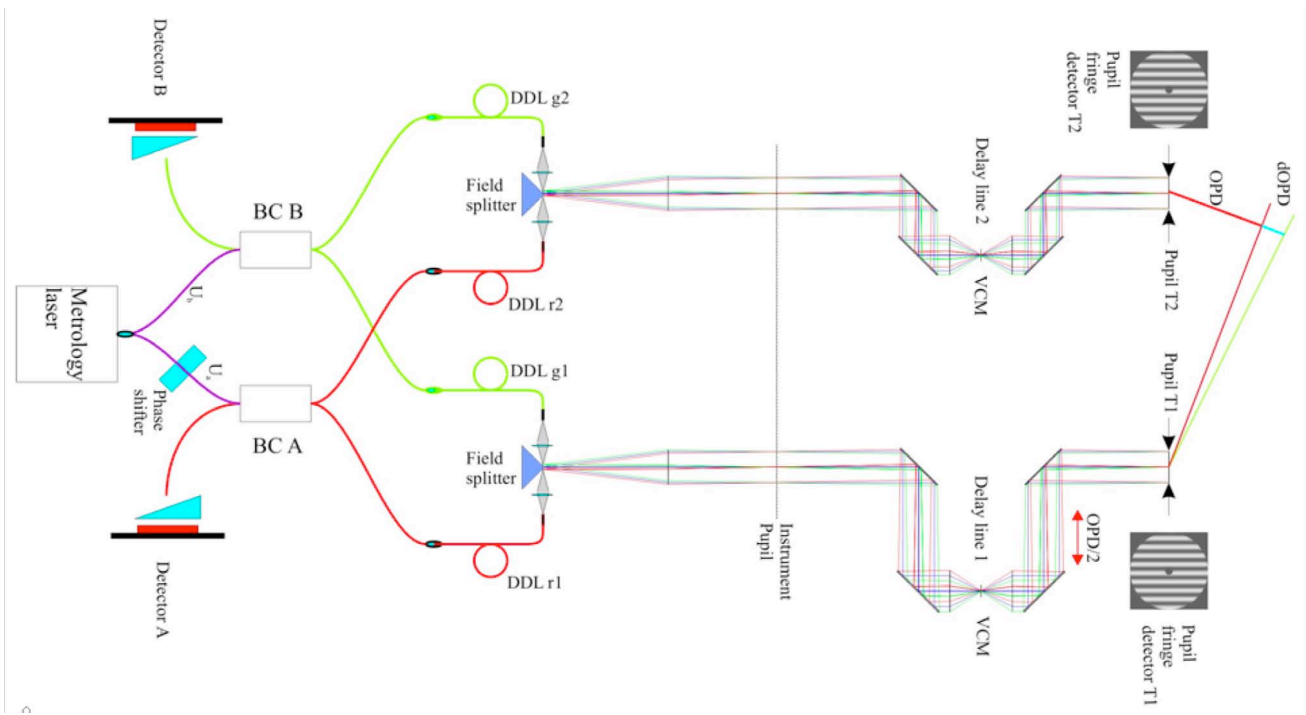

Figure 20: Working principle of the novel metrology concept foreseen for GRAVITY. Two laser beams are injected into the beam combiners, traveling towards the telescopes and forming a fringe pattern in the entrance pupil plane. Monitoring the latter allows tracking the dOPD.

## 6.9 Metrology

The laser metrology measures the differential optical path between the two stars introduced by the VLTI beam relay and the beam combiner instrument. It measures the differential optical path between the two stars for each telescope. This allows measuring the full optical path between the telescope M2 and the beam combiner, and to trace the full pupil (figure 20). The metrology laser is injected at the exit of the integrated optics beam-combiner. The interference pattern is observed in scattered light on M2 and on dedicated scatterers on the M2 spider arms (i.e. in M1 space) with infrared cameras at the UTs. At the ATs, the detection will use the direct light sensed with photodiodes mounted at the M3 spider arms. The metrology is integrated as a phase-shifting interferometer and uses a high-power (2 W), high-stability (30 MHz), linearly polarized cw-laser at 1908 nm. The laser is fed into the integrated optics via single-mode polarization maintaining fibers, fiber-based beam splitters, electric phase-shifters and classical coupling optics. The coupling efficiency to the integrated optics is optimized via piezo-electric x,y,z motor stages [43]. The principle of detecting fringes in scattered light from M2 has been tested successfully in May 2007 at one UT.

## 6.10 Data Reduction software

The data reduction software provides all tasks for the calibration and reduction of the data collected with the instrument. These are the online instrument level routines to monitor the instrument during operation, the classical pipeline, which runs in Paranal in parallel to the observation to assess the quality of the obtained data, the quality control and stability pipeline to output the instrument monitoring parameters, the routines to produce uncalibrated science products, i.e. visibilities and phases, and the routines for calibrated data products from the observation of a reference source. The GRAVITY software architecture is original compared to other competing instruments, due to its industrial-like architecture, integration in the ESO data flow, and public-oriented accessibility. The implementation of an image reconstruction software is still to be done.

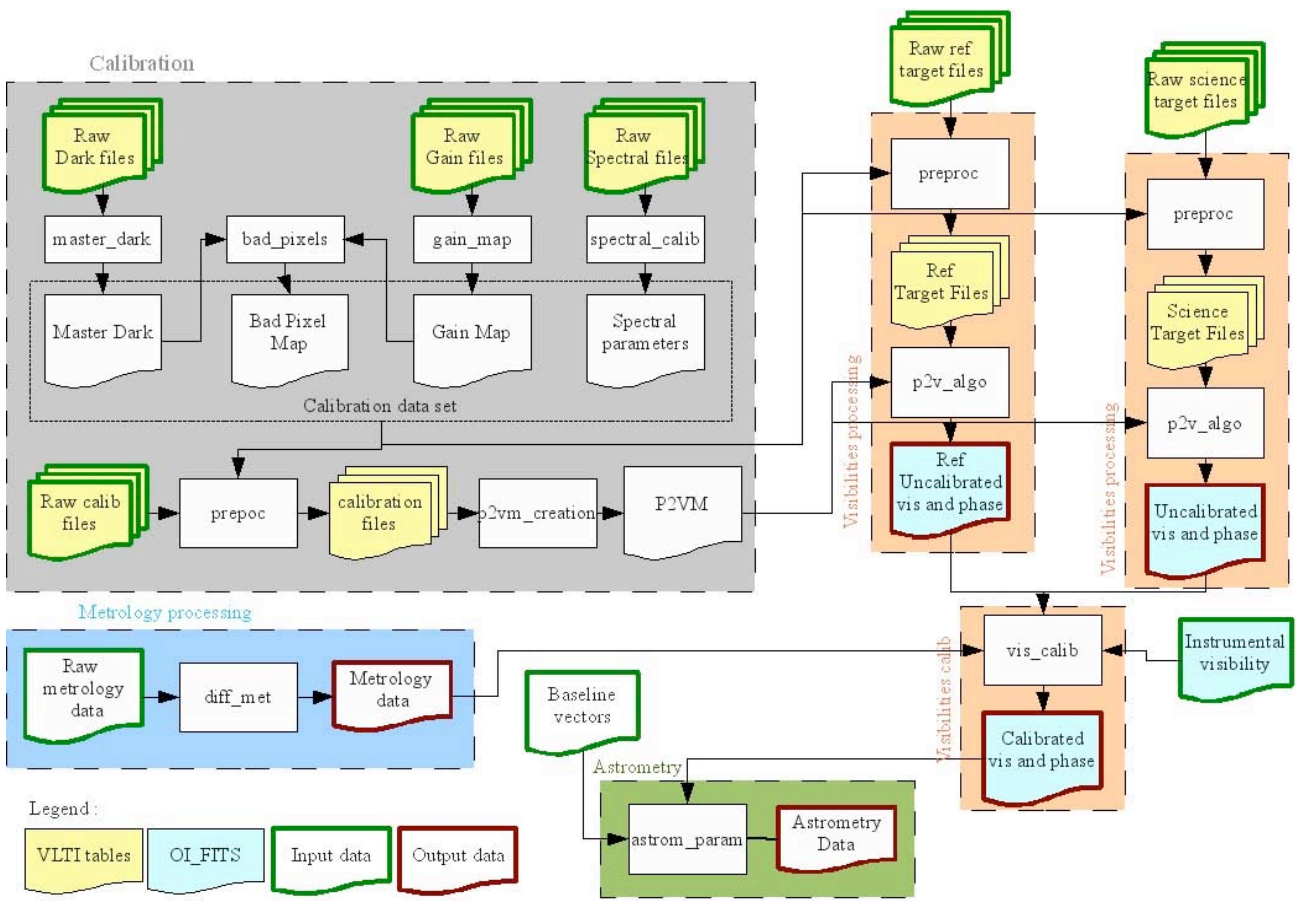

Figure 21: Overview of the preliminary design of the data reduction software.

# 7. EXPECTED PERFORMANCE

We have developed both an analytic performance system analysis using average quantities, as well as an explicit simulation of the expected performance. These two approaches complement each other and consistently show that GRAVITY will deliver the desired performance.

The GRAVITY fringe-tracker will work reliably for median seeing conditions on a $m_K = 10$ star, and provide a residual rms OPD of less than 250 nm on the science beam combiner when using a 6-electron RON detector (as the SELEX device). Implementing a detector with 13-electron RON detector yields 400 nm residual OPD level, limited by flux dropouts. The current handling of these dropouts is not yet optimized; but an improved algorithm could push the fringe-tracking performance even with the higher RON detector to the level of 250 nm. The final level of residual OPD fluctuations will in addition strongly depend on the vibrations induced by the UT telescopes. From our simulations, we conclude that the vibration induced OPD fluctuations should be lowered to below 200 nm rms [44].

The phase noise per baseline on a $m_K = 15$ star in a 5-minute observations will be ~ 4 nm. Including atmospheric residuals and systematic errors the resulting narrow angle astrometric accuracy will be of the order 10 μas. The signal to noise ratio for the visibility of a $m_K = 16$ star in a 100 s observing time will be ~ 14, the respective phase error ~ 0.04 rad. The dynamic range in the reconstructed image will be > 3 mag. Size and position measurements will be possible for objects as faint as $m_K > 19$ in one night for isolated, single sources with a bright phase reference within 2". Table 2 summarizes the interferomteric performance estimates. The astrometric accuracy achievable in a 5-minute observation of a $m_K=15$ target with a $m_K = 10$ reference star is about 12 μas. Table 3 shows the according break-down of the astrometric error budget.

|  | **Expected performance** | **Specification** |
|---|---|---|
| **Fringe tracking on $m_K = 10$ star (correct handling of flux-dropouts)** | ~ 250 nm rms OPD on science channel | < 300 nm (goal 200 nm) |
| **Astrometry between a $m_K = 10$ primary and $m_K = 15$ secondary star** | ~ 12 μas in 5 minutes | < 30 μas (goal 10 μas) in 5 minutes |
| **Interferometric imaging on $m_K = 16$ in 100 s, with a $m_K = 10$ phase reference** | S/N Visibility ~ 14  σ(Phase) = 0.04 rad  Dynamic range > 3 mag | S/N Visibility > 10  σ(Phase) < 0.1 rad  - |
| **Size and position measurements of an isolated, single source with a bright phase reference within 2"** | $m_K > 19$ in 6 hours | - |

Table 2: Expected interferometric performance and specification for GRAVITY. For these numbers it is assumed that the level of mechanical vibrations in the VLTI will be reduced such that the corresponding residual OPD is at the level of ~ 200 nm rms, and that consequently integration times of minutes are possible.

| Error term | Error value | δOPD error 1" separation 83 m baseline | |
|---|---|---|---|
| NAB determination | 0.5 mm | 2.42 nm | 6.01 μas |
| BC phase measurements ($T_{INT}$ = 5min) | $\lambda_s$: 1.2 nm (R = 1800) $\Phi_{FT}$: 0.8 nm $\Phi_{SC}$: 4.0 nm | 4.0 nm | 10.0 μas |
| Metrology | $\lambda_m$: 0.36 pm (R = $10^6$) $\Phi_M$: 1 nm | 1.41 nm | 3.50 μas |
| Dispersion | $n^{\lambda m}/n^{\lambda s}$: $10^{-6}$ (after calibration) DL hysteresis: 36 nm | 0.40 nm | 1.00 μas |
| Lateral pupil error | Centering: 15 mm Tip-Tilt: 15 mas | 3.08 nm | 7.66 μas |
| Telescope pointing error | 10 arcsec | 2.26 nm | 5.62 μas |
| Atmospheric lensing | NAB: 25 μm | 0.12 nm | 0.30 μas |
| Relativistic and geographic effect | | 1 nm | 2.48 μas |
| Anisoplanatism (TINT= 5min) | | 5.46 nm | 13.56 μas |
| Star azimuthal position | 75 mas | 0.23 nm | 0.57 μas |
| **Sub-Total** | | **8.3 nm** | **20.7 μas** |
| **TOTAL (using 6 baselines)** | **Factor 1/√3** | **4.8 nm** | **12.0 μas** |

Table 3: Astrometric error budget for GRAVITY